\documentclass[letterpaper, 10 pt, conference]{ieeeconf}

\IEEEoverridecommandlockouts
\usepackage{graphics} 
\usepackage{epsfig} 
\usepackage{times} 

\usepackage{algorithm}
\usepackage{algpseudocode}
\usepackage{amsmath,amsfonts,amssymb,amsthm}
\usepackage{chngcntr}
\usepackage{xparse}
\usepackage{graphicx}
\usepackage{makecell}
\usepackage{accents}
\usepackage{subcaption}
\usepackage{xcolor}

\theoremstyle{plain}
\newtheorem{assumption}{Assumption}
\newtheorem{lemma}{Lemma}
\newtheorem{proposition}{Proposition}

\title{\LARGE \bf
Computing Forward Reachable Sets for Nonlinear Adaptive Multirotor Controllers
}

\author{Juyeop Han and Han-Lim Choi
\thanks{This research was supported by Unmanned Vehicles Core Technology Research and Development Program through the National Research  Foundation of Korea(NRF), Unmanned Vehicle Advanced Research Center(UVARC) funded by the Ministry of Science and ICT, the Republic of Korea (2020M3C1C1A01082375)}
\thanks{Juyeop Han and Han-Lim Choi are with the KAIST Institutes for Robotics
        and the Department of Aerospace Engineering,
        Korea Advanced Institute of Science and Technology (KAIST),
        Daejeon, 34141, South Korea.
        {\tt\small jyhan@lics.kaist.ac.kr, hanlimc@kaist.ac.kr}}%
}

\begin{document}

\maketitle
\thispagestyle{empty}
\pagestyle{empty}

\begin{abstract}

In multirotor systems, guaranteeing safety while considering unknown disturbances is essential for robust trajectory planning. The Forward reachable set (FRS), the set of feasible states subject to bounded disturbances, can be utilized to identify robust and collision-free trajectories by checking the intersections with obstacles. However, in many cases, the FRS is not calculated in real time and is too conservative to be used in actual applications. In this paper, we address these issues by introducing a nonlinear disturbance observer (NDOB) and an adaptive controller to the multirotor system. We express the FRS of the closed-loop multirotor system with an adaptive controller in augmented state space using Hamilton-Jacobi reachability analysis. Then, we derive a closed-form expression that over-approximates the FRS as an ellipsoid, allowing for real-time computation. By compensating for disturbances with the adaptive controller, our over-approximated FRS can be smaller than other ellipsoidal over-approximations. Numerical examples validate the computational efficiency and the smaller scale of our proposed FRS.

\end{abstract}

\section{INTRODUCTION}

Recently, multirotors have attracted attention due to their agile maneuvers, simple dynamics, and potential for complex tasks, such as surveillance \cite{Alexis2009} and aerial manipulation \cite{Kim2013}. Mostly, safety in the multirotor is referred to as avoiding collision with various obstacles and is indispensable to prevent catastrophic accidents. \cite{Mellinger2011},  \cite{Mueller2015} exploit convex optimization and motion primitives to generate a feasible and collision-free trajectory for a multirotor. However, these approaches do not fully guarantee safety as a result of unexpected factors, such as air drags, wind gusts, and model uncertainty, which cause disturbances in the actual system dynamics.


Therefore, safety should be ensured by considering disturbances when planning the trajectory of a multirotor. The forward reachable set (FRS) is one of the most popular concepts for analyzing the safety of the system dynamics. A FRS is a set of all states produced by all possible bounded disturbances given an initial state set. 

The Hamilton-Jacobi (HJ) reachability analysis is a method for computing reachable sets based on an optimal control theory \cite{Mitchell2005}. The HJ partial differential equation (PDE) for the sub-level set is derived by the principle of dynamic programming. However, the HJ reachability analysis is associated with the "curse of dimensionality;" that is, solving the HJ PDE requires computation exponential to the number of states of the system \cite{Bansal2017}. To resolve and mitigate this issue, the decomposition of the system dynamics \cite{Chen2018} and a deep-learning-based approach \cite{Bansal2021} can be applied.
Many studies without using the HJ reachability analysis provides a method by which to determine the outer or inner approximation of an exact FRS of a linear system to well-known geometric shapes, such as a zonotope \cite{Girard2005} or an ellipsoid \cite{Kurzhanski2000, Chernousko2004}, as it is intractable to attempt to find the exact FRSs of many high-dimensional nonlinear systems.

Safety-ensured trajectory planning against disturbances can be performed by combining existing planning approaches and the FRS. \cite{Majumdar2017} computes the FRSs of motion primitives, referred to as a funnel, using sums of squares (SOS) programming and utilizes these funnels for trajectory planning. \cite{Kim2018} designs a multirotor controller compensating for aerodynamic effects and computes tight FRSs by SOS programming as well. \cite{Herbert2017, Fridovich2018} convert a robust trajectory planning problem into a pursuit-evasion game. The maximum tracking error is computed via a HJ reachability analysis and is utilized to avoid obstacles. All of these studies require an offline phase for the FRS computation. On the other hand, \cite{Seo2019, Seo2020} compute the FRSs of the nominal trajectories of the multirotor in real time. These studies approximate the FRS to an ellipsoid in an analytical expression by exploiting a generalized Hopf formula \cite{Lions1986}. Although \cite{Seo2019, Seo2020} compute the high dimensional approximation of the FRS in real time unlike \cite{Girard2005,Kurzhanski2000, Chernousko2004}, the approximated FRS is too conservative to be applied to agile trajectory planning.

Our research aims to compute small FRSs in a receding horizon using a nonlinear disturbance observer (NDOB) and a nonlinear adaptive controller for a multirotor in real time. To achieve this objective, we extend earlier work \cite{Seo2019}, \cite{Seo2020} by applying the adaptive controller as a feedback controller with the assumptions that the disturbance is differentiable and that its time derivative is bounded. The NDOB and the adaptive controller for the multirotor are presented and the stability is also proven to justify their use. Tight future disturbance bounds are predicted by the NDOB for the computation of the smaller FRSs. Moreover, we augment the estimated disturbance and true disturbance to the closed-loop dynamics of the multirotor. The FRSs of the augmented dynamics are formulated by a HJ reachability analysis and are approximated to ellipsoids. Finally, the ellipsoidal FRSs are projected onto the original state space in an analytical expression.

The main contributions of the paper are as follows:

\begin{itemize}
    \item We propose a mathematical formulation for the FRSs of the multirotor system combined with the adaptive controller and their ellipsoidal over-approximation in the augmented state space.
\end{itemize}
\begin{itemize}
    \item We implement real-time computation of the approximated FRSs which are much smaller than a baseline \cite{Seo2020}.
\end{itemize}

\subsection{Notation}
\label{Notation}
In this paper, most vectors are described in bold, but their $i$-th element is not described in bold and is denoted as $(\cdot)_i$. $\mathbf{0}_{n\times m} \in \mathbb{R}^{n\times m}$ and $I_n \in \mathbb{R}^{n\times n}$ are correspondingly a $n \times m$ zero matrix and a $n \times n$ identity matrix. $\Vert \cdot \Vert$ and $\Vert \cdot \Vert_\infty$ likewise represent a second norm and a $\infty$-norm.  $\textrm{Proj}_{\mathbb{A}}(\cdot)$ and $\textrm{Prop}_{\mathbb{A}}(\cdot)$ are the set projection and propagation operation to the space $\mathbb{A}$. $\mathbb{E}(s, K) = \{ x \in \mathbb{R}^n \vert \; (x - s)^TK(x - s) \leq 1 \}$ is an ellipsoid for a vector $s \in \mathbb{R}^n$ and a positive definite matrix $K \in \mathbb{R}^{n \times n}$. Finally, $\oplus$ stands for the Minkowski sum between two sets, and $\bigoplus$ represents a series of the Minkowski sums of multiple sets.

\section{PROBLEM FORMULATION}
\label{PROBLEM FORMULATION}
This section introduces the dynamics of a multirotor (\ref{subsec:II-A}) and formulates its forward reachable set (FRS) with the adaptive controller (\ref{subsec:II-B}). The augmented system of the multirotor combined with the adaptive controller is also introduced, followed by a linearization of the augmented system. The FRS is reformulated based on the linearization (\ref{subsec:II-C}).

\subsection{Multirotor System Dynamics} \label{subsec:II-A}
We denote the simplified state and control input of the multirotor dynamics as $\mathbf{x} = [\mathbf{p}^T \; \mathbf{v}^T \; \mathbf{\Phi}^T]^T \in \mathbb{X} \subset \mathbb{R}^9$ and $\mathbf{u} = [F \; \mathbf{\mathbf{\Omega}}^T]^T \in \mathbb{U} \subset \mathbb{R}^4$. The position, the velocity and the Euler angle are represented as $\mathbf{p} \in \mathbb{R}^3$, $\mathbf{v} \in \mathbb{R}^3$ and $\mathbf{\Phi} \in \mathbb{R}^3$, respectively. $F \in \mathbb{R}$ and $\mathbf{\mathbf{\Omega}} \in \mathbb{R}^3$ are likewise the normalized thrust and the angular velocity of the multirotor. We consider the simplified multirotor dynamics with disturbance $\mathbf{d} \in \mathbb{R}^3$ introduced by external forces as a control-affine form:
\begin{equation}
    \label{eqn:control affine dynamics of multirotor}
    \dot{\mathbf{x}} = f(\mathbf{x}) + g_1(\mathbf{x})\mathbf{u} + g_2\mathbf{d}
\end{equation}
with
\begin{equation*}
\begin{split}
    & f(\mathbf{x}) = [\mathbf{v}^T \; (-g\mathbf{e}_3)^T \; \mathbf{0}_{1\times3}]^T \\
    & g_1(\mathbf{x}) = \left[
    \begin{matrix}
    \mathbf{0}_{1 \times 3} & (R(\Phi)\mathbf{e}_3)^T & \mathbf{0}_{1 \times 3} \\
    \mathbf{0}_{3 \times 3} & \mathbf{0}_{3 \times 3}  & C(\Phi)^T\\ \end{matrix} \right]^T \\
    & g_2 = [\mathbf{0}_{3\times3} \; I_3 \; \mathbf{0}_{3\times3}]^T
\end{split}
\end{equation*}
denoting the magnitude of gravity acceleration $g \in \mathbb{R}$, and $\mathbf{e}_3 = [0 \; 0 \; 1]^T \in \mathbb{R}^3$. $R(\Phi) \in \mathbb{R}^{3 \times 3}$ is the rotation matrix from the body frame to the inertial frame. $C(\Phi)$ is the mapping matrix from the angular velocity to the Euler angle rate.

\subsection{Forward Reachable Set for Multirotor with Adaptive Controller} \label{subsec:II-B}
In this research, the multirotor is controlled by an adaptive controller to track a reference trajectory. The adaptive controller compensates for the disturbance estimated by a nonlinear disturbance observer (NDOB). An assumption is required for the NDOB to estimate the disturbance:
\begin{assumption}
\label{as:1}
The disturbance vector and its time derivative are bounded by positive real constants in each channel. That is, the disturbance at time $t$ is an element of
    \begin{equation*} \label{eqn:disturbance set}
        \begin{split}
            \mathcal{D}(t) = \{ \mathbf{d}(t) \vert \forall \tau \in [t_0, t],\, \exists \mathbf{w}(\tau) \in \mathcal{W}, \, \dot{\mathbf{d}}(\tau) = \mathbf{w}(\tau),& \\
            {\vert d_i(\tau) \vert} \leq L_i ,\, \mathbf{d}(t_0) \in \mathcal{D}(t_0) &\}
        \end{split}
    \end{equation*}
  with $\mathcal{W} = \{\mathbf{w} \vert \; {\vert w_i \vert} \leq \beta_i \} \subset \mathbb{R}^3$, and  constant vectors, $\mathbf{L} \in \mathbb{R}^3$ and $\beta \in \mathbb{R}^3$.
\end{assumption}
We also consider the NDOB as the dynamics of the estimated disturbance $\hat{\mathbf{d}} \in \mathbb{R}^3$: 
\begin{equation}
    \label{eqn:NDOB equation}
    \dot{\hat{\mathbf{d}}} = \zeta(\hat{\mathbf{d}}, \mathbf{d})
\end{equation}
The exact formulation of the NDOB will be described in Section \ref{subsec:III-A}, and the dynamics of the estimated disturbance will be shown to be linear.

Subsequently, we build an adaptive controller to track a differential flat output, $\mathbf{r}(t) = [\mathbf{p}_{\mathbf{r}}(t)^T \; \psi_{\mathbf{r}}(t)]^T \in \mathbb{R}^4$, where $\mathbf{p}_{\mathbf{r}}(t) \in \mathbb{R}^3$ (resp. $\psi_{\mathbf{r}}(t) \in \mathbb{R}$) is the reference position and yaw angle \cite{Mellinger2011}. The closed-loop dynamics combined with the adaptive controller can be expressed as
\begin{equation} \label{eqn:closed-loop dynamics}
    \begin{split}
        \mathbf{u}(t) &= \kappa(\mathbf{x}(t), \hat{\mathbf{d}}(t) \,; \mathbf{r}(t)) \\
        \dot{\mathbf{x}}(t) &= h(\mathbf{x}(t), \hat{\mathbf{d}}(t) \, ; \mathbf{r}(t)) + g_2\mathbf{d}(t)
    \end{split}
\end{equation}
where $\kappa(\mathbf{x}(t), \hat{\mathbf{d}}(t) \,; \mathbf{r}(t))$ is the adaptive controller, and $ h(\mathbf{x}(t), \hat{\mathbf{d}}(t) \, ; \mathbf{r}(t)) = f(\mathbf{x}(t)) + g_1(\mathbf{x}(t))\kappa(\mathbf{x}(t), \hat{\mathbf{d}}(t) \,; \mathbf{r}(t))$. The exact formulation of $\kappa$ will be described in Section \ref{subsec:III-A} as well.

Let us define the FRS of the closed-loop dynamics at time $t$ as follows:
\begin{equation} \label{eqn:multirotor FRS}
    \begin{split}
        \mathcal{X}(t) = \{& \mathbf{x}(t) \vert \, \forall \tau \in [t_0, t],\, \exists  \mathbf{d}(\tau) \in \mathcal{D}(\tau), \\
    &\dot{\mathbf{x}}(\tau) = h(\mathbf{x}(\tau), \hat{\mathbf{d}}(\tau) ; \mathbf{r}(\tau)) + g_2\mathbf{d}(\tau), \\ 
    &\dot{\hat{\mathbf{d}}}(\tau) = \zeta(\hat{\mathbf{d}}(\tau), \mathbf{d}(\tau)), \\ 
    & \mathbf{x}(t_0) \in \mathcal{X}(t_0), \, \hat{\mathbf{d}}(t_0) \in \hat{\mathcal{D}}_0 \}
    \end{split}
\end{equation}
where $\hat{\mathcal{D}}_0$ is the initial set of the estimated disturbance.

The objective of the research is to compute of the ellipsoidal over-approximation of the FRS, $\mathcal{X}(\tau)$, in receding horizon, $ \tau \in [t_0,  t_f] $ in real time.

\subsection{Augmented System and Linearization} \label{subsec:II-C}
We augment the mulitrotor system (\ref{eqn:control affine dynamics of multirotor}) by adding the estimated disturbance $\mathbf{\hat{d}}$ and the real disturbance $\mathbf{d}$ to the state $\mathbf{x}$ since the NDOB (\ref{eqn:NDOB equation}) and the \textit{Assumption \ref{as:1}} must be considered to compute the FRS (\ref{eqn:multirotor FRS}). Let $ \mathbf{y} = [\mathbf{x}^T \; \hat{\mathbf{d}}^T \; \mathbf{d}^T]^T \in \mathbb{Y} \subset \mathbb{R}^{15}$ be the augmented state of the multirotor. From (\ref{eqn:disturbance set}), (\ref{eqn:NDOB equation}) and (\ref{eqn:closed-loop dynamics}), the dynamics of the augmented state $\mathbf{y}$ is derived as
\begin{equation} \label{eqn:augmented dynamics}
    \begin{split}
        \dot{\mathbf{y}}(t) &=
        \left[ \begin{matrix}
        h(\mathbf{x}(t), \hat{\mathbf{d}}(t) \, ; \mathbf{r}(t)) + g_2\mathbf{d}(t) \\
        \zeta(\hat{\mathbf{d}}(t), \mathbf{d}(t)) \\
        \textbf{0}_{3 \times 1} \\
        \end{matrix} \right]
        + F\mathbf{w}(t) \\
        &= \xi(\mathbf{y}(t) ; \mathbf{r}(t)) + F\mathbf{w}(t)
    \end{split}
\end{equation}
with $F = [\mathbf{0}_{3 \times 12} \; I_3]^T \in \mathbb{R}^{15 \times 3}$. The augmented dynamics (\ref{eqn:augmented dynamics}) is linearized to apply the generalized Hopf formula \cite{Lions1986} in Section \ref{ADAPTIVE FORWARD REACHABLE SET COMPUTATION}-B to compute the FRS of the augmented dynamics:
\begin{equation} \label{eqn:linearized dynamics}
    \dot{\mathbf{e}}_\mathbf{y}(t) = A(t)\mathbf{e_y}(t) + F\mathbf{w}(t) + \mu(\mathbf{y}(t), \mathbf{w}(t); \mathbf{r}(t))
\end{equation}

with the error state $\mathbf{e_y}(t) = \mathbf{y}(t) - \mathbf{y_r}(t)$, the state matrix $A(t) = \left.{\partial \xi / \partial \mathbf{y}}\right|_{(\mathbf{y_r}(t); \mathbf{r}(t))}$. The vector of linearization error is represented as $\mu(\mathbf{y}(t), \mathbf{w}(t); \mathbf{r}(t)) = [\Delta_{\mathbf{p}}^T \; \Delta_{\mathbf{v}}^T \; \Delta_{\mathbf{\Phi}}^T \; \mathbf{0}_{6 \times 1}^T]^T$ where $\Delta_{\mathbf{p}}$, $\Delta_{\mathbf{v}} $, and $\Delta_{\mathbf{\Phi}} \in \mathbb{R}^3$ denote the linearization errors in position, velocity, and Euler angle space respectively. Note that the NDOB dynamics will be shown to be linear in Section \ref{subsec:III-A}. Additionally, $\mathbf{y_r}(t) \in \mathbb{Y}$ is a zero-disturbance reference state with initial condition $\mathbf{y_r}(t_0) = \mathbf{y}(t_0)$ such that $\forall \tau \in (t_0, t]$, $\mathbf{d_r}(\tau) = \mathbf{0}_{3 \times 1}$ and $\dot{\mathbf{y}}_\mathbf{r}(\tau) = \xi(\mathbf{y_r}(\tau);\mathbf{r}(\tau))$.

\begin{assumption}
    \label{as:2}
    The linearization error vector $\mu(\mathbf{y}(t), \mathbf{w}(t); \mathbf{r}(t))$ is either bounded or neglected. In the case that it is bounded, each channel of $\mu(\mathbf{y}(t), \mathbf{w}(t); \mathbf{r}(t))$ is constrained within $\mathbf{M_y} = [\mathbf{M_p}^T \; \mathbf{M_v}^T \; \mathbf{M_\Phi}^T \; \mathbf{0}_{6 \times 1}^T]^T \in \mathbb{R}^{15}$ with maximum errors in the position, velocity, and the Euler angle space, $\mathbf{M_p} \in \mathbb{R}^3, \; \mathbf{M_v} \in \mathbb{R}^3, \; \mathbf{M_\Phi} \in \mathbb{R}^3$, respectively. That is,
    \begin{equation*}
        \vert \mu(\mathbf{y}(t), \mathbf{w}(t); \mathbf{r}(t))_i \vert \leq M_{y,i}
    \end{equation*}
\end{assumption}
According to \textit{Assumption 2}, the error dynamics (\ref{eqn:linearized dynamics}) becomes
\begin{equation} \label{eqn:modified linearized dynamics}
    \dot{\mathbf{e}}_\mathbf{y}(t) = A(t)\mathbf{e_y}(t) + \Bar{F}\Bar{\mathbf{w}}(t)
\end{equation}
with $\Bar{\mathbf{w}} = [\Delta_\mathbf{p}^T \; \Delta_\mathbf{v}^T \; \Delta_{_\mathbf{\Phi}}^T \; \mathbf{w}^T]^T \in \mathbb{R}^{12}$ and $\Bar{F} = [F' \; F] \in \mathbb{R}^{15 \times 12}$ (resp. $\Bar{\mathbf{w}} = \mathbf{w}$ and $\Bar{F} = F$) with $F' = [I_9 \; \mathbf{0}_{9 \times 6}]^T \in \mathbb{R}^{15 \times 9}$ when the linearized error is considered and bounded (resp. neglected). Let $n_{\Bar{w}}$ be a dimension of $\Bar{w}$. The vector $\Bar{\mathbf{w}}$ is bounded to the vector $\Bar{\beta} = [\mathbf{M_p}^T \; \mathbf{M_v}^T \; \mathbf{M_\Phi}^T \; \beta^T]^T$ (resp. $\Bar{\beta} = \beta$) $\in \mathbb{R}^{n_{\Bar{w}}}$ when the linearization error is considered and bounded (resp. neglected). $\Bar{\mathcal{W}}$ is defined as a set of all possible vectors $\Bar{\mathbf{w}}$. As a result, the vector $\Bar{\mathbf{w}}$ can be shown as the bounded disturbance of (\ref{eqn:modified linearized dynamics}).

The FRS of the linearized error dynamics (\ref{eqn:modified linearized dynamics}) without considering the possible disturbance set $\mathcal{D}(t)$ is defined as shown below.
\begin{equation} \label{eqn:FRS of linearized Dynamics}
    \begin{split}
        \mathcal{E}_{\mathcal{Y}}(t) = &\{ \mathbf{e_y}(t) \vert \, \forall \tau \in [t_0, t], \,
        \exists \mathbf{w}(\tau) \in \mathcal{W}(\tau), \\
        & \dot{\mathbf{e}}_\mathbf{y}(\tau) = A(\tau)\mathbf{e_y}(\tau) + \Bar{F}\Bar{\mathbf{w}}(\tau), \, \mathbf{e_y}(t_0) \in \mathcal{E}_{\mathcal{Y}}(t_0) \}
    \end{split}
\end{equation}
Since $\mathcal{E}_{\mathcal{Y}}(t) \cap \textrm{Prop}_{\mathbb{Y}}(\mathcal{D}(t))$  represents the FRS of the error dynamics considering the disturbance set $\mathcal{D}(t)$, the FRS of the augmented dynamics (\ref{eqn:augmented dynamics}) is the same as $\mathbf{y_r}(t) + \mathcal{E}_{\mathcal{Y}}(t) \cap \textrm{Prop}_{\mathbb{Y}}(\mathcal{D}(t))$, which is a translation of the FRS of the augmented error dynamics with the augmented reference state $\mathbf{y_r}(t)$.

According to the definition of $\mathcal{X}$ (\ref{eqn:multirotor FRS}) and $\mathcal{E}_{\mathcal{Y}}$ (\ref{eqn:FRS of linearized Dynamics}), the following relationship is satisfied if the projection of $\mathcal{E}_{\mathcal{Y}}(t_0) \cap \textrm{Prop}_{\mathbb{Y}}(\mathcal{D}(t_0))$ to the original state space $\mathbb{X}$ and the projection to the estimated disturbance space are correspondingly identical to $\mathcal{X}(t_0)$ and $\hat{\mathcal{D}}_0$:
\begin{equation}\label{eqn:FRS projection}
    \mathcal{X}(t) = \mathbf{x_r}(t) + \textrm{Proj}_{\mathbb{X}}(\mathcal{E}_{\mathcal{Y}}(t) \cap \textrm{Prop}_{\mathbb{Y}}(\mathcal{D}(t)))
\end{equation}
with $\mathbf{x_r}(t) = \textrm{Proj}_{\mathbb{X}}(\mathbf{y_r}(t))$. consequently, our goal is to over-approximate the right-hand side term of (\ref{eqn:FRS projection}) as an ellipsoid.

\section{NONLINEAR DISTURBANCE OBSERVER AND ADAPTIVE CONTROLLER}
\label{NONLINEAR DISTURBANCE OBSERVER AND ADAPTIVE CONTROLLER}

\subsection{Nonlinear Disturbance Observer and Adaptive Controller Design}
\label{subsec:III-A}
The structure of the NDOB  for the multirotor dynamics (\ref{eqn:control affine dynamics of multirotor}) introduces an internal state vector $\rho \in \mathbb{R}^3$ to estimate the true disturbance $d$ \cite{Mohammadi2017}:
\begin{equation} \label{eqn:NDOB}
    \begin{gathered}
        \hat{\mathbf{d}} = \rho
        + p(\mathbf{x}) \\
        \dot{\rho} = -L_d(\mathbf{x})g_2\rho - L_d(\mathbf{x})[f(\mathbf{x}) + g_1(\mathbf{x})\mathbf{u} + g_2p(\mathbf{x})]
    \end{gathered}
\end{equation}
with $p(\mathbf{x}) = \alpha_{\mathbf{d}}\mathbf{v} \in \mathbb{R}^3$ and $L_d(\mathbf{x}) = \partial p / \partial \mathbf{x} = \alpha_{\mathbf{d}}g_2^T \in \mathbb{R}^{3 \times 9}$. $\alpha_{\mathbf{d}} \in \mathbb{R}$ is a positive real constant. Let $\mathbf{e}_d = \mathbf{d} - \hat{\mathbf{d}}$ be the disturbance error. By differentiating $\mathbf{\hat{d}}$ in (\ref{eqn:NDOB}), the estimated disturbance dynamics (\ref{eqn:NDOB equation}) is derived as $\zeta(\mathbf{\hat{d}}, \mathbf{d}) =\alpha_{\mathbf{d}} \mathbf{e}_{\mathbf{d}}$.

At this point, we design the adaptive controller $\kappa$ to compensate for the estimated disturbance $\hat{\mathbf{d}}$ in a manner similar to that in \cite{Kim2018}. Let $\mathbf{e_p} = \mathbf{p} - \mathbf{p_r}$ and $\mathbf{e_v} = \mathbf{v} - \dot{\mathbf{p}}_\mathbf{r}$ be the position error and the velocity error, respectively. The thrust of the adaptive controller is expressed as
\begin{equation} \label{eqn: thrust of adaptive controller}
    \begin{split}
        & \mathbf{f}_d = -k_{\mathbf{p}}\mathbf{e_p} -k_{\mathbf{v}}\mathbf{e_v} + g\mathbf{e}_3 + \ddot{\mathbf{p}}_\mathbf{r} - \hat{\mathbf{d}} \\
        & \mathbf{z}_{b,d} = \frac{\mathbf{f}_d}{\Vert \mathbf{f}_d \Vert},\quad f = \mathbf{f}_d^T\mathbf{z}_b = {\Vert \mathbf{f}_d \Vert} \mathbf{z}_{b,d}^T\mathbf{z}_b
    \end{split}
\end{equation}
where $k_{\mathbf{p}} \in \mathbb{R}$ and $k_{\mathbf{v}} \in \mathbb{R}$ are the positive feedback gains and $\mathbf{z_b}$ is the thrust direction of the multirotor. The desired roll $\phi_\mathbf{r}(t)$ and pitch $\theta_\mathbf{r}(t)$ can be computed from the thrust (\ref{eqn: thrust of adaptive controller}) and the flat output $\mathbf{r}(t)$ \cite{Mellinger2011}. The angular velocity $\mathbf{\Omega}$ is controlled to follow the reference Euler angle $\Phi_r(t) \in \mathbb{R}^3$ according to  the attitude controller having exponential stability, such as the geometric controller \cite{Lee2010}, \cite{Lee2010-1}. 

Suppose that $\phi_b$ is the angle between the desired thrust direction $\mathbf{z}_{b,d}$ and the true thrust direction $\mathbf{z}_{b}$. Because the controller makes the attitude error exponentially stable, the angle $\phi_b \in \mathbb{R}$ is bounded so that there exists sufficiently small $s_m$ such that $\vert \sin{\phi_b} \vert \leq s_m$ if the initial attitude is close to the initial reference attitude.

\subsection{Uniformly Ultimate Boundedness}

The dynamics of the position error and the velocity error of the closed-loop system and the disturbance error dynamics of the NDOB are derived as

\begin{equation}
\begin{split}
\label{eqn:error dynamics}
    \dot{\mathbf{e}}_\mathbf{p} &= \mathbf{e_v} \\
    \dot{\mathbf{e}}_\mathbf{v} &= -k_{\mathbf{p}}\mathbf{e_p} -k_{\mathbf{v}}\mathbf{e_v} + \mathbf{e_d} + ({\Vert \mathbf{f}_d \Vert}\sin{\phi_b})\mathbf{l} \\
    \dot{\mathbf{e}}_\mathbf{d} &= \mathbf{w} - \alpha_{\mathbf{d}}\mathbf{e_d}
\end{split}
\end{equation}

where $\mathbf{l} \in \mathbb{R}^3$ is the unit vector of $(\mathbf{z_b}\cdot\mathbf{z_{b,d}})\mathbf{z_b} - \mathbf{z_{b,d}}$.

We show that the error of (\ref{eqn:error dynamics}) is uniformly ultimately bounded (UUB) to verify the convergence of the presented adaptive controller.
\begin{lemma} \label{prop:UUB_disturbance}
    The disturbance error dynamics ${e}_d$ is UUB.
\end{lemma}
\textit{proof}: See Appendix A. \qed

The rate and radius of convergence in \emph{Lemma \ref{prop:UUB_disturbance}} will be used for tight disturbance bound prediction in Section \ref{ADAPTIVE FORWARD REACHABLE SET COMPUTATION}-B.

\begin{proposition}
    \label{prop:translation error UUB}
    The dynamics of the position error ${e}_p$ and the velocity error ${e}_v$ are UUB if $k_{\mathbf{v}} > 1$, $\alpha_{\mathbf{d}} > \frac{1}{4}(\frac{1}{k_{\mathbf{p}}} + \frac{1}{k_{\mathbf{v}} - 1})$ and $s_m < -\lambda_{min}(Q_1)/\lambda_{min}(Q_2)$, where $\lambda_{min}(Q_1)$ and $\lambda_{min}(Q_2)$ are the corresponding minimum eigenvalues of $Q_1$ and $Q_2$, and
    \begin{equation*}
        \begin{gathered}
        Q_1 = \left[\begin{matrix}
            k_{\mathbf{p}} & 0 & -\frac{1}{2} \\
            0 & k_{\mathbf{v}}-1 & -\frac{1}{2} \\
            -\frac{1}{2} & -\frac{1}{2} & \alpha_{\mathbf{d}} \end{matrix} \right], \\
        Q_2 = \left[\begin{matrix}
            -k_{\mathbf{p}} & -\frac{1}{2}(k_{\mathbf{p}} + k_{\mathbf{v}}) & -\frac{1}{2} \\
            -\frac{1}{2}(k_{\mathbf{p}} + k_{\mathbf{v}}) & -k_{\mathbf{v}} & -\frac{1}{2} \\
            -\frac{1}{2} & -\frac{1}{2} & 0 \end{matrix} \right]
        \end{gathered}
    \end{equation*}
\end{proposition}
\textit{proof}: See Appendix B. \qed 

\subsection{Disturbance Bound Prediction on the Prediction Horizon}

Suppose that the initially estimated disturbance is set to $\hat{\mathbf{d}}(0) = \mathbf{0}_{3 \times 1}$ and that the disturbance $\mathbf{d}$ is observed during time interval $[0, t_0]$ by the NDOB (\ref{eqn:NDOB}). Let future possible disturbance predictions start from time $t_0$ for the FRS computation. We derive the $\infty$-norm bound of the disturbance error at time $t_0$, ${\Vert \mathbf{e_d}(t_0) \Vert}_\infty \leq r(t_0)$, from the proof of the \emph{Lemma \ref{prop:UUB_disturbance}} with $r(t) = \max ( {\Vert \mathbf{L} \Vert} \exp{[-(1-\theta_1)\alpha_{\mathbf{d}}t]}, {\Vert \beta \Vert} /(\theta_1\alpha_{\mathbf{d}}) )$.

Given that the disturbances and their time derivatives are bounded to the constant vector radius $L$ and $m$ according to the \emph{Assumption 1}, for the future possible disturbance set, the following inclusive relationship for all $\tau \in [t_0, t]$ is satisfied:
\begin{equation*} \label{eqn:error bound propagation}
\begin{gathered}
        \mathcal{D}(\tau) \subset 
    {\mathcal{D}_{\mathcal{F}}}(\tau) \cap  {\mathcal{D}_{\mathcal{B}}} \\
    \text{s.t. } {\mathcal{D}_{\mathcal{B}}} = \{\mathbf{d} \vert \vert d_i \vert \leq L_i \} \\
     {\mathcal{D}_{\mathcal{F}}}(\tau)  = \{ \mathbf{d}(\tau) \vert \; \vert d_i(\tau) - \hat{d}_i(t_0) \vert \leq r(t_0) + \beta_i(\tau - t_0) \}
\end{gathered}
\end{equation*}

We set the upper and lower bounds of $i$-th channel of ${\mathcal{D}_{\mathcal{F}}}(\tau) \cap  {\mathcal{D}_{\mathcal{B}}}$ to $d_{L,i}(\tau)$ and $d_{U,i}(\tau)$. Let the center and the edge of the future disturbance be $\mathbf{d}_m(\tau) \in \mathbb{R}^3$ and $\mathbf{d}_M(\tau) \in \mathbb{R}^3$, where the $i$-th elements of these vectors are $d_{m,i}(\tau) = (d_{L,i}(\tau) + d_{U,i}(\tau))/2$ and $d_{M,i}(\tau) = (d_{U,i}(\tau) -d_{L,i}(\tau))/2$, respectively. As a result, the future possible disturbances at time $t$ are expressd as
\begin{equation*} \label{eqn:disturbance constraint}
    \mathcal{D}_P(t) = \big\{\mathbf{d}(t) \vert \, \forall \tau \in [t_0, t], {\vert d_i(\tau) - d_{m,i}(\tau) \vert} < d_{M,i}(\tau) \big\}
\end{equation*}
and, the relationship, $\textrm{Prop}_{\mathbb{Y}}(\mathcal{D}(t)) \subset \textrm{Prop}_{\mathbb{Y}}(\mathcal{D}_P(t))$, is satisfied.
\section{ADAPTIVE FORWARD REACHABLE SET COMPUTATION}
\label{ADAPTIVE FORWARD REACHABLE SET COMPUTATION}

\subsection{Coordinate Transformation and Hamilton-Jacobi Reachability Analysis}

We aim to approximate $\mathbf{x_r}(t) + \mathcal{E}_{\mathcal{Y}}(t) \cap \textrm{Prop}_{\mathbb{Y}}(\mathcal{D}(t))$ to an ellipsoid in a way similar to that in \cite{Seo2019}, \cite{Seo2020}. Suppose that  $\mathcal{E}_{\mathcal{Y}}(t)$ is the sub-zero level set of the value function $V(\mathbf{e_y}, t) \in \mathbb{R}$:
\begin{equation*}
    \label{FRS value function expression}
    \mathcal{E}_{\mathcal{Y}}(t)
    = \{\mathbf{e_y} \vert \, V(\mathbf{e_y}, t) \leq 0\} \;
    \textrm{s.t.} \; \mathcal{E}_{\mathcal{Y}}(t_0) = \{\mathbf{s} \vert l_0(\mathbf{s}) \leq 0\}
\end{equation*}
where $l_0(\mathbf{s}) \in \mathbb{R}$ is the initial convex value function. Let $l_0(\mathbf{s}) = (\mathbf{s}-\mathbf{q_0})^TQ_0^{-1}(\mathbf{s}-\mathbf{q_0}) - 1$ so that $\mathcal{E}_{\mathcal{Y}}(t_0) = \mathbb{E}(\mathbf{q_0}, Q_0^{-1})$ for vector $\mathbf{q_0} \in \mathbb{R}^{15}$ and positive definite matrix $Q_0 \in \mathbb{R}^{15 \times 15}$.

By multiplying the inverse of the state transition matrix $\Psi(t)$ by the error dynamics (\ref{eqn:modified linearized dynamics}), the system is transformed to a new dynamics for the variable $\eta(t) = \Psi^{-1}(t)\mathbf{e_y}(t)$ :
\begin{equation}
    \label{eqn:tranformed dynamics}
    \dot{\eta}(t) = \Psi^{-1}(t)\Bar{F}\Bar{\mathbf{w}}(t) = {\Bar{F}}_{\eta}(t)\Bar{\mathbf{w}}(t).
\end{equation}

The HJ reachability analysis \cite{Bansal2017} provides the following HJ partial differential equation to find the value function $V_{\eta}(\eta, t)$:
\begin{equation} \label{eqn: HJ PDE2}
    \begin{split}
        &\frac{\partial V_{\eta}}{\partial t} + H_{\eta}(\nabla V_{\eta},t)  = 0 \\
        &V_{\eta}(\eta,t_0) = l_{0,\eta}(\eta) \\
        &H_{\eta}(\mathbf{p},t) = \underset{\Bar{w}(t)\in \Bar{\mathcal{W}}}{\textrm{max}}\mathbf{p}^T ({\Bar{F}}_{\eta}(t)\Bar{\mathbf{w}}(t))
    \end{split}
\end{equation}
where $(\cdot)_{\eta}$ denotes the corresponding formulation of $(\cdot)$ in $\eta$-space.
The following explicit solution of (\ref{eqn: HJ PDE2}) is found by the generalized Hopf formula:
\begin{equation} \label{eqn:FRS eta Set}
    \begin{gathered}
    \mathcal{E}_{\eta}(t) = \mathcal{E}_{\mathcal{Y}}(t_0) \oplus \mathcal{G}_{\eta}(t) \\
    \textrm{s.t.} \; \mathcal{G}_{\eta}(t) = \bigoplus_{i=1}^{n_{\Bar{w}}}\mathcal{G}_{\eta,i}(t) \\
    \mathcal{G}_{\eta,i}(t) = \Big\{ {\Bar{\beta}}_i \int_{t_0}^{t}{\Bar{F}}_{\eta,i}(\tau) \textrm{sign}(Q_0^{-\frac{1}{2}} {\Bar{F}}_{\eta,i}(\tau)^T \nu)d\tau \Big\vert {\Vert \nu \Vert}^2 \leq 1 \Big\}
    \end{gathered}
\end{equation}
where ${\Bar{F}}_{\eta,i}(t)$ is the $i$-th column of ${\Bar{F}}_{\eta}(t)$.
For more detailed information, we refer the reader to \cite{Seo2019}, \cite{Seo2020}.

\subsection{Ellipsoidal Approximation}

Set $\mathcal{G}_{\eta}(t)$ is approximated to the ellipsoid $\mathbb{E}(\mathbf{0}_{15 \times 1}, B_{\eta}^{-1}(t))$ \cite{Seo2019}, \cite{Seo2020}, as shown below.
\begin{equation} \label{eqn:Ellipsoidal Approximation}
    \begin{gathered}
    \mathcal{G}_{\eta}(t) \subset \mathbb{E}(\mathbf{0}_{15 \times 1}, B_{\eta}^{-1}(t)) \\
    \textrm{s.t.} \; B_{\eta}(t) = \sum^{n_{\Bar{w}}}_{i=1} \frac{B_{\eta,i}(t)}{a_i} \\
    B_{\eta,i}(t) = 
    (t-t_0)\int_{t_0}^{t}({\Bar{\beta}}_i^2 {\Bar{F}}_{\eta,i}(\tau){\Bar{F}}_{\eta,i}(\tau)^T + \epsilon I_{15})\,d\tau
    \end{gathered}
\end{equation}
where $a_i = \sqrt{\text{tr}(B_{\eta,i}(t))} / \big( \sum^{n_{\Bar{w}}}_{i=1}\sqrt{\text{tr}(B_{\eta,i}(t))}  \, \big)$ is a coefficient that enables $B_{\eta}(t)$ to be a minimal-trace containing the sum of ellipsoids \cite{Durieu1996} and $\epsilon$ is any positive constant.

According to (\ref{eqn:FRS eta Set}) and (\ref{eqn:Ellipsoidal Approximation}), a conservative ellipsoidal approximation of $\mathcal{E}_{\eta}(t)$ satisfies 
\begin{equation} \label{eqn:epsilon_eta approximation}
    \begin{gathered}
    \mathcal{E}_{\eta}(t) \subset \mathbb{E}(\mathbf{q_0}, Q_\eta^{-1}(t)) \\
    \textrm{s.t. } Q_\eta(t) = Q_0/a_1 +B_{\eta}(t)/a_2
    \end{gathered}
\end{equation}
where the coefficients, $a_1$ and $a_2$ are designed for ellipsoidal fusion with the minimal trace as (\ref{eqn:Ellipsoidal Approximation}). Next,  $\mathbb{E}(\mathbf{q_0}, Q_\eta^{-1}(t)))$ is converted to the approximation of $\mathcal{E}_{\mathcal{Y}}(t)$ in $\mathbf{e_y}$-space:

\begin{equation} \label{eqn:epsilon_y approximation}
    \begin{gathered}
    \mathcal{E}_{\mathcal{Y}}(t) \subset \mathbb{E}(\mathbf{q_y}, Q_y^{-1}(t)) \\
    \textrm{s.t.} \; Q_y(t) = \Psi(t)Q_\eta(t)\Psi(t)^T, \, \mathbf{q_y} = \Psi(t)\mathbf{q_0}
    \end{gathered}
\end{equation}

We approximate $\mathcal{D}_P(t)$ to a conservative ellipsoid, $\mathbb{E}(\mathbf{d}_m(t), \Lambda(t))$ with $\Lambda = \text{diag}([\lambda_1, \lambda_2, \lambda_3])$. For $\Lambda^{-1}(t)$ to have a minimal trace, a convex problem is shown below and solved by applying the KKT condition \cite{Boyd2004}.
\begin{equation*}
    \underset{\lambda_i}{\min} \sum_{i=1}^3 \frac{1}{\lambda_i} \text{ s.t. } \sum_{i=1}^3 \lambda_i d_{M,i}^2 \leq 1
\end{equation*}
The optimal matrix having a minimal inverse trace is 
\begin{equation} \label{eqn: Lambda_star}
    \Lambda^* = \textrm{diag}([\lambda_1^*, \lambda_2^*, \lambda_3^*]) \textrm{ s.t. } \lambda_i^* = \big(\sum_{j=1}^3 d_{M,j} \big)^{-1}d_{M,i}^{-1}.    
\end{equation}

As a result, we can express $\mathcal{E}_{\mathcal{Y}}(t) \cap \textrm{Prop}_{\mathbb{Y}}(\mathcal{D}_P(t))$ as a conservative approximation of the intersections of ellipsoids:
\begin{multline} \label{eqn:ellipsoid intersection}
    \mathcal{E}_{\mathcal{Y}}(t) \cap \textrm{Prop}_{\mathbb{Y}}(\mathcal{D}_P(t))
    \subset \\
    \mathbb{E}(\mathbf{q_y}(t), Q_y^{-1}(t)) \cap \text{Prop}_{\mathbb{Y}}(\mathbb{E}(\mathbf{d}_m(t), \Lambda^*(t)))
\end{multline}
Note that the propagation of the ellipsoid to higher dimension space can be represented in the form of an ellipsoid by adding zeros to the elements of its center and matrix corresponding to the propagated dimensions. Let $\mathbf{d}_m^\mathbb{Y}(t) \in \mathbb{R}^{15}$ (resp. $\Lambda^\mathbb{Y}(t) \in \mathbb{R}^{15 \times 15}$) be the center (resp. matrix) of $\text{Prop}_{\mathbb{Y}}(\mathbb{E}(\mathbf{d}_m(t), \Lambda^*(t)))$.

We again approximate (\ref{eqn:ellipsoid intersection}) to a single ellipsoid \cite{Durieu1996}:
\begin{equation} \label{eqn:solution S}
    \begin{gathered}
    S(t) = \mathbb{E}(\mathbf{c}(t), M(t)) \\
    \textrm{s.t.} \;
    N(t) = b Q_y^{-1}(t) + (1-b)\Lambda^\mathbb{Y}(t), \\
    \delta = (1-b)(\mathbf{d}_m^\mathbb{Y}(t))^T\Lambda^\mathbb{Y}(t)\mathbf{d}_m^\mathbb{Y}(t) - \mathbf{c}^TN\mathbf{c},\\
    \mathbf{c}(t) = N^{-1}(bQ_y^{-1}(t)\mathbf{q_y}(t) (1-b)\Lambda^\mathbb{Y}(t)d_m^\mathbb{Y}(t)), \\
    M(t) = \frac{N(t)}{1 - \delta}
    \end{gathered}
\end{equation}
with $0 \leq b \leq 1$. Note that the ellipsoid $S(t)$ will be utilized as the initial ellipsoid $\mathbb{E}(\mathbf{q_0}, Q_0^{-1})$ to calculate the propagated $S(t + \Delta t)$ after time step $\Delta t$.

Finally, the approximation of the FRS (\ref{eqn:FRS projection}) is derived from the projection of (\ref{eqn:solution S}) to the original state space $\mathbb{X}$ as $\mathbf{x_r}(t) + \textrm{Proj}_{\mathbb{X}}(S(t))$. $\textrm{Proj}_{\mathbb{X}}(S(t))$ is an ellipsoid \cite{Ros2002} and can be represented as
\begin{equation} \label{eqn:projection S}
   \mathbb{E}(\textrm{Proj}_{\mathbb{X}}(\mathbf{c}(t)), \,M_{11}(t) - M_{12}(t)M_{22}^{-1}(t)M_{21}(t)),
\end{equation}
where $M_{11}(t) \in \mathbb{R}^{9 \times 9}$, $M_{12}(t) \in \mathbb{R}^{9 \times 6}$, $M_{21}(t) \in \mathbb{R}^{6 \times 9}$ and $M_{22}(t) \in \mathbb{R}^{6 \times 6}$ are block components of $M(t)$ divided by the original state $x$ and the disturbance state $[\hat{\mathbf{d}}^T \; \mathbf{d}^T]^T$.

The computation process for over-approximation of FRS during receding time horizon, $T_{rec}$, is showed in \textit{Algorithm \ref{alg:1}}. The approximated FRS is computed at every discretized time step, $t_i \in T_{rec}$.

\begin{algorithm} 
\caption{Over-approximation of FRS as ellipsoid} \label{alg:1}
\begin{algorithmic}[1]
\Statex \textbf{Input:} $T_{rec} = \{t_0, t_1, \cdots ,t_k\}$, $\{ A(t_i) \}_{i=0}^{k-1}$, $\Bar{F}$, $\Bar{\beta}$, $(q_0(t_0), Q_0(t_0))$, $\{ ( d_m(t_i), d_M(t_i) ) \}_{i=0}^{k-1}$, $\{ x_r(t_i) \}_{i=0}^{k-1}$
\State $i \rightarrow 0$
\While{$i \leq k-1$}
    \State \textbf{Over-approximation of $\mathcal{E}_{\mathcal{Y}}$:}
    \State Assign $(\exp({A(t_i)(t_{i+1}-t_i)}) \rightarrow \Psi(t_i)$
    \State Assign $\Psi^{-1}(t_i)\Bar{F} \rightarrow \Bar{F_{\eta}}$
    \State Compute $B_{\eta}(t_i)$ using $\Bar{F_{\eta}}$ and $\Bar{\beta}$ (\ref{eqn:Ellipsoidal Approximation})
    \State Compute $Q_{\eta}(t_i)$ using $Q_0(t_i)$ and $B_{\eta}(t_i)$ (\ref{eqn:epsilon_eta approximation})
    \State Assign $(\Psi(t_i)\mathbf{q_0}(t_i), \, \Psi(t_i)Q_{\eta}(t_i)\Psi(t_i)^T) \rightarrow (\mathbf{q}_y(t_i), Q_y(t_i)) $ (\ref{eqn:epsilon_y approximation})
     
    \State \textbf{Over-approximation of $\mathcal{E}_{\mathcal{Y}} \cap \text{Prop}_{Y}(\mathcal{D}(t))$:}
    \State Compute $\Lambda^*(t_i)$ using $(\mathbf{d}_m(t_i), \mathbf{d}_M(t_i))$ (\ref{eqn: Lambda_star})
    \State Compute $c(t_i)$ and $M(t_i)$ of $S(t_i)$ using ($\mathbf{q}_y(t_i)$, $Q_y(t_i)$, $\mathbf{d}_m(t_i)$, $\Lambda^*(t_i)$) (\ref{eqn:solution S})
    \State Assign $(c(t_i), M(t_i)) \rightarrow (\mathbf{q_0}(t_{i+1}), Q_0(t_{i+1}))$
    
    \State \textbf{Over-approximation of $\mathcal{X}(t_i)$:}
    \State Compute $\mathbf{x}_r(t_i) + \text{Proj}_{\mathbb{X}}(S(t_i))$ (\ref{eqn:projection S})
    \State $i+1  \rightarrow i$
\EndWhile
\State \textbf{return} $\{\mathbf{x}_r(t_i) + \text{Proj}_{\mathbb{X}}(S(t_i))\}_{i=0}^{k-1}$
\end{algorithmic}
\end{algorithm}

\section{NUMERICAL EXAMPLES}
\label{RESULTS}

We implement two numerical scenarios, ’scenario 1’ and ’scenario 2’, to verify our method for FRS computations on a desktop computer. The objective of these scenarios is to demonstrate that the FRS computed by our method has a smaller trace and volume compared to that computed by the baseline \cite{Seo2020}. The controller used in both scenarios is identical to the one introduced in Section \ref{NONLINEAR DISTURBANCE OBSERVER AND ADAPTIVE CONTROLLER}, except that in the baseline, the estimated disturbance ˆd is always set to zero.

\subsection{Setup and Scenarios}

The laptop has an AMD Ryzen 7 6800H with a 3.2GHz base clock and 16GB of RAM. The computation method is programmed in MATLAB R2020b on Linux 18.04. The reference trajectory in all examples is a circular trajectory rotated with a 30$^\circ$ roll and a 30$^\circ$ yaw. The center of the trajectory is $[0\;0\;0]^T$. The radius of the trajectory is 10m and 0.6 rad/s, in the two cases. We compute the FRSs with a 2.7s prediction horizon. The parameters used in the two numerical examples are shown in Table \ref{tab:parameters}. We set the disturbance bound L to describe the wind blowing from the lateral direction. $K$ denotes a vector composed of feedback gains including the position and velocity feedback gain, $k_{\mathbf{p}}$ and $k_{\mathbf{v}}$ in Section \ref{NONLINEAR DISTURBANCE OBSERVER AND ADAPTIVE CONTROLLER}. $\mathbf{q_0}(t_0)$ and $Q_0(t_0)$ denote the center of an ellipsoid and the ellipsoidal matrix representing the initial set at the initial time, $t_0$, of the prediction horizon, respectively.

Scenario 1 shows that the FRSs were computed in the first prediction horizon by the proposed method with or without considering the linearization error and the baseline \cite{Seo2020}. Here, 500 trajectories starting from the same points were generated with random disturbances satisfying \emph{Assumption \ref{as:1}}. In scenario 2, multirotors with the two different controllers used to the baseline and the proposed method followed the reference trajectory within 15.2s. The FRSs computed by the proposed method considering the linearization error and those by \cite{Seo2020} were computed every 2.5s. The starting positions were identical to the reference starting position in all scenarios. Additionally, we added a 10m$\times$10m$\times$20m cuboid obstacle centered at $[0\;0\;0]^T$ to visually compare the FRSs generated by the two methods.

\begin{table}[t]
    \centering
    \caption{Parameters for Numerical Examples}
    \begin{tabular}{|c|c|} \hline
    \textbf{Parameter} & \textbf{Value}\\
        \hline\hline
        $\alpha_{\mathbf{d}}$ & $ 2.0 \textrm{/s}^2 $ \\ \hline
        $\theta_1$ & 0.8 \\ \hline
        $\Delta t$ & 0.02s \\ \hline
        b & 0.99 \\ \hline
        $K$ & $[18.0\;6.0\;7.0\;21.0]^T$ \\ \hline
        $\mathbf{L}$ & $ [3.0 \textrm{m/s}^2 \; 3.0\textrm{m/s}^2 \; 1.0 \textrm{m/s}^2] $ \\ \hline
        $\beta$ & $ [2.0 \textrm{m/s}^3\;2.0 \textrm{m/s}^3\;2.0 \textrm{m/s}^3] $ \\ \hline
        $\mathbf{M_p}$ & [0.001m/s \; 0.001m/s \; 0.001m/s]$^T$ \\ \hline
        $\mathbf{M_v}$ & [0.01m/s$^2$ \; 0.01m/s$^2$ \; 0.01m/s$^2$]$^T$ \\ \hline
        $\mathbf{M_\Phi}$ & [0.01rad/s \; 0.01rad/s \; 0.01rad/s]$^T$ \\ \hline
        $\mathbf{q_0}(t_0)$ & \makecell{[0m\;0m\;0m\;0m/s\;0m/s\;0m/s\;0rad\;0rad\;0rad\\
        0$\textrm{m/s}^2$\;0$\textrm{m/s}^2$\;0$\textrm{m/s}^2$ \;$d_{m,1}(t_0)$\;$d_{m,2}(t_0)$\;$d_{m,3}(t_0)$]$^T$} \\ \hline
        $Q_0(t_0)$ &
        \makecell{diag([0.05m, 0.05m, 0.05m, 0.05$\textrm{m/s}$,
        0.05$\textrm{m/s}$, 0.05$\textrm{m/s}$, \\
        0.05rad, 0.05rad, 0.05rad,
        0.05$\textrm{m/s}^2$, 0.05$\textrm{m/s}^2$, 0.05$\textrm{m/s}^2$, \\
        $3d_{M,1}(t_0)$, $3d_{M,2}(t_0)$, $3d_{M,3}(t_0)$])$^2$} \\\hline
        
    \end{tabular}
    \label{tab:parameters}
\end{table}

\subsection{Results}

\begin{figure}[htb!]
  \begin{subfigure}{.5\textwidth}
      \centering
      \includegraphics[width=\linewidth]{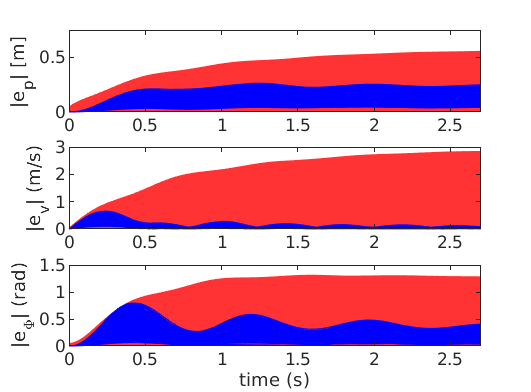}
      \caption{Baseline}
      \label{fig:sfig1}
    \end{subfigure}

    \begin{subfigure}{.5\textwidth}
      \centering
      \includegraphics[width=\linewidth]{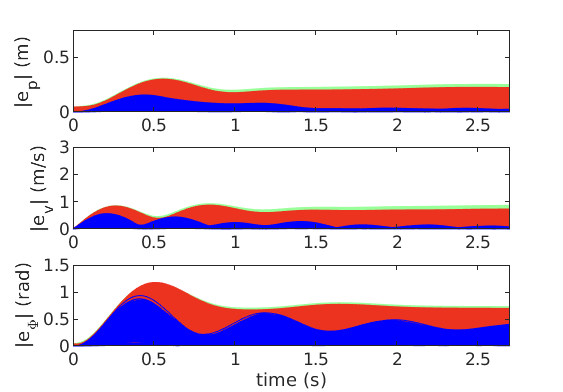}
      \caption{Proposed method}
      \label{fig:sfig2}
    \end{subfigure}

  \caption{ The maximum error bounds and actual errors obtained from baseline \cite{Seo2020} (a) and our method (b) are presented. The possible error bounds neglecting linearization errors (red shaded areas). The additional possible error bounds considering linearization errors (green shaded area). The magnitude of actual state errors (blue area).}
  \label{fig:error}
\end{figure}

\begin{figure}[htb!]
    \centering
    \includegraphics[width = .5\textwidth]{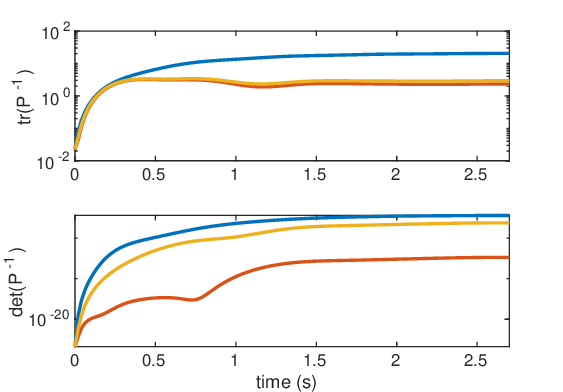}
    \caption{The traces (top) and determinants (down) of the inverse matrices for the ellipsoidal approximation of FRSs. The baseline \cite{Seo2020} (blue lines). Proposed method neglecting linearization errors (brown lines). Proposed method considering linearization errors (yellow lines).}
    \label{fig:trance and det}
\end{figure}

\begin{figure}[htb!]
    \begin{subfigure}{.24\textwidth}
      \centering
      \includegraphics[width=\linewidth]{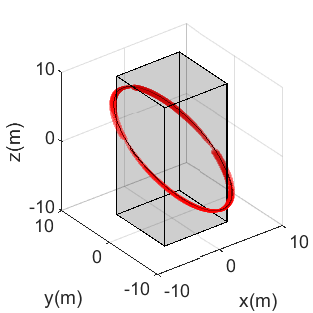}
      \caption{Baseline (side view)}
      \label{fig:sfig5}
    \end{subfigure}
    \hfill
    \begin{subfigure}{.24\textwidth}
      \centering
      \includegraphics[width=\linewidth]{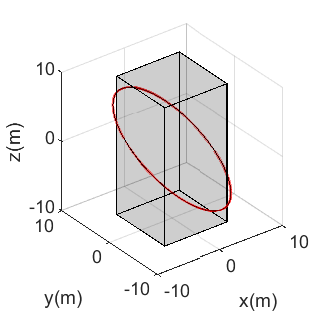}
      \caption{Proposed method (side view)}
      \label{fig:sfig6}
    \end{subfigure}

    \begin{subfigure}{.24\textwidth}
      \centering
      \includegraphics[width=\linewidth]{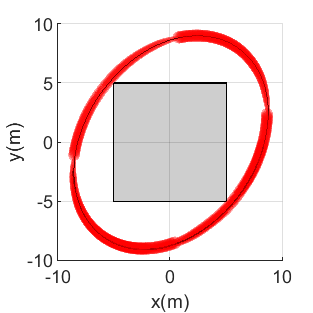}
      \caption{Baseline (top view)}
      \label{fig:sfig7}
    \end{subfigure}
    \hfill
    \begin{subfigure}{.24\textwidth}
      \centering
      \includegraphics[width=\linewidth]{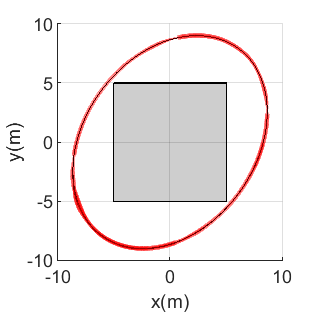}
      \caption{Proposed method (top view)}
      \label{fig:sfig8}
    \end{subfigure}

  \caption{The side and top views of the cuboid obstacles (gray boxes), FRSs in the position space (red shaded areas) and the trajectory following the circular reference trajectory (black lines) for the baseline \cite{Seo2020} and proposed method. }\label{fig:FRS visualization}
\end{figure}

Fig. \ref{fig:error} illustrates the possible error bounds obtained from the FRSs and the sampled trajectory errors in the position, velocity, and Euler angle in the first scenario. Fig. \ref{fig:trance and det} describes the trace and the determinant of the inverse matrices of the ellipsoidal FRSs in Scenario 1 as Fig. \ref{fig:error}. Fig. \ref{fig:FRS visualization} visualizes the FRSs in the position space and trajectories following the reference trajectories with the two controllers in Scenario 2.

Fig. \ref{fig:error} shows that all sampled trajectories belong to the possible error bound generated by the baseline \cite{Seo2020} and the proposed method with or without considering linearization errors. All figures verify that the FRS calculated by our method is much smaller than the baseline FRS on a log scale. Moreover, Fig. \ref{fig:FRS visualization} shows that FRSs generated by the baseline intersect with the edges of the cuboid obstacle, while the FRSs generated by the proposed method do not. In Scenario 2, average computation times for propagation of FRSs of the baseline and of the proposed method when neglecting or considering linearization errors are 13.1ms ($\pm$1.22ms), 20.7ms ($\pm$1.34ms) and 26.7ms ($\pm$1.71ms), respectively. Although the computation time of the proposed method is slightly slower than that of the baseline \cite{Seo2020}, our method is fast enough for real-time trajectory planning considering disturbances.

In conclusion, the results verify that the FRS computed by the proposed method is much smaller than the baseline FRS and is implemented in real time.

\section{CONCLUSION}
\label{CONCLUSION}

In this paper, we proposed a real-time computation method for a small forward reachable set (FRS) to guarantee the safety of multirotor dynamics with disturbances. We introduce an nonlinear disturbance observer (NDOB) and an adaptive controller into the system dynamics and over-approximate the FRS to an ellipsoid in an analytical expression. Numerical examples verify that the proposed method rapidly computes the smaller FRS of the multirotor system than the baseline \cite{Seo2020} does. Also, our approach can be extended to other mobile system dynamics, such as ground vehicles. One limitation is that the approximated FRS is only available when the linearization error of the system is negligible or when the error is bounded and known. Considering the smaller FRS computed by the proposed method than the baseline, the authors expect that the proposed method will be highly applicable to plan agile and robust trajectories in real time.


\bibliographystyle{unsrt}
\bibliography{reference}

\appendix
\subsection{Proof of Lemma 1}
Suppose that the function $V_1 = \frac{1}{2}\mathbf{e_d}^T\mathbf{e_d}$. The time derivative of $V_1$ derived by (\ref{eqn:error dynamics}) is $ \dot{V}_1 =  - \alpha_{\mathbf{d}} \mathbf{e_d}^T\mathbf{e_d} + \mathbf{w}^T\mathbf{e_d} $. The inequality, ${\Vert \mathbf{w} \Vert} \leq {\Vert \beta \Vert}$, is satisfied by the \emph{Assumption 1} so that $\dot{V}_1$ is developed as
\begin{equation} \label{disturance Lyapunov diff2}
    \begin{split} 
    \dot{V}_1
    &\leq  - \alpha_{\mathbf{d}}(1-\theta_1) { \Vert \mathbf{e_d}  \Vert}^2 - \alpha_{\mathbf{d}}\theta_1 {\Vert \mathbf{e_d} \Vert}^2 + {\Vert \beta \Vert} {\Vert \mathbf{e_d} \Vert} \\
    & \leq - 2\alpha_{\mathbf{d}}(1-\theta_1) V_1,
    \quad \forall \Vert \mathbf{e_d} \Vert \geq \frac{{\Vert \beta \Vert}}{\alpha_{\mathbf{d}}\theta_1}.
    \end{split}
\end{equation}
with $0 < \theta_1 < 1$.

The disturbance error ${e}_d$ converges to a sphere centered to the origin with a radius of $ {\Vert \beta \Vert} / \alpha_{\mathbf{d}}\theta_1 $ with exponential rate $\alpha_{\mathbf{d}}(1-\theta_1)$ \cite{Khalil2002}.

\subsection{Proof of Proposition 1}
The proof is similar to the proof for the UUB of the controller presented in \cite{Kim2018}. Suppose the sum of the functions $V = V_1 + V_2$ such that $V_1$ is the function in the \emph{Lemma \ref{prop:UUB_disturbance}} and $V_2 = \frac{1}{2}\mathbf{e_t}^T P \mathbf{e_t}$ with $e_t = [\mathbf{e_p}^T \; \mathbf{e_v}^T]^T \in \mathbb{R}^6$ and
\begin{equation}
    P = \left[
\begin{matrix}
(k_{\mathbf{p}} + k_{\mathbf{v}})I_3 & I_3\\
I_3 & I_3 \\ \end{matrix} \right]. \nonumber
\end{equation}

From the desired thrust (\ref{eqn: thrust of adaptive controller}) and bounded angle $\phi_b$, we derive
\begin{equation}
    \label{eqn:RHS sub}
    {\Vert \mathbf{f}_d \Vert}{\Vert\sin{\phi_b}\mathbf{w}\Vert} \leq s_m(k_{\mathbf{p}}{\Vert \mathbf{e_p} \Vert} + k_{\mathbf{v}}{\Vert \mathbf{e_v} \Vert} + M + {\Vert \mathbf{e_d} \Vert} )
\end{equation}
with $M \in \mathbb{R}^+$ such that ${\Vert g\mathbf{e}_3 + \ddot{\mathbf{p}}_\mathbf{r} + \mathbf{d} \Vert} \leq M$ when $\ddot{p}_r$ does not diverge to infinity.

The following inequality related to the time derivative of $V$ is developed by considering (\ref{eqn:error dynamics}) and (\ref{eqn:RHS sub}):
\begin{equation}
    \begin{split}
    \label{eqn:System UUB1}
    \dot{V} \leq& -\mathbf{e}^TQ_1\mathbf{e} \\
    & + {\Vert \mathbf{f}_d \Vert}{\Vert\sin{\phi_b}\mathbf{w}\Vert}({\Vert \mathbf{e_p} \Vert} + {\Vert \mathbf{e_v} \Vert}) + {\Vert \beta \Vert}{\Vert \mathbf{e_d} \Vert} \\
    \leq& - \mathbf{e}^TQ\mathbf{e} + N{\Vert \mathbf{e} \Vert}    
    \end{split}
\end{equation}
with $\mathbf{e} = [{\Vert \mathbf{e_p} \Vert} \; {\Vert \mathbf{e_v} \Vert} \; {\Vert \mathbf{e_d} \Vert}]^T \in \mathbb{R}^3$, $N = \sqrt{2s_m^2M^2 + {\Vert \beta \Vert}^2}$ and $Q = Q_1 + s_mQ_2$. The first and second conditions of the \emph{Proposition \ref{prop:translation error UUB}}, $k_{\mathbf{v}} > 1$ and $\alpha_{\mathbf{d}} > \frac{1}{4}(\frac{1}{k_{\mathbf{p}}} + \frac{1}{k_{\mathbf{v}} - 1})$, ensure that $Q_1$ is positive definite. The last condition, $s_m < -\lambda_{min}(Q_1)/\lambda_{min}(Q_2)$, ensures that $Q$ is positive definite despite the fact that $Q_2$ is not positive definite.

Via  a method similar to (\ref{disturance Lyapunov diff2}), the inequality (\ref{eqn:System UUB1}) becomes
\begin{equation} \label{System UUB2}
    \begin{split}
        \dot{V} \leq & -\lambda_{min}(Q){\Vert \mathbf{e} \Vert}^2 + N{\Vert \mathbf{e} \Vert}  \\
        \leq & -\lambda_{min}(Q)(1 - \theta_2){\Vert \mathbf{e} \Vert}^2, \,  \forall{\Vert \mathbf{e} \Vert} \geq \frac{N}{\lambda_{min}(Q) \theta_2}
    \end{split}
\end{equation}
with $0 < \theta_2 < 1$.

The error dynamics (\ref{eqn:error dynamics}) are UUB.
\end{document}